\documentclass{jpsj3}
\usepackage{txfonts}
\usepackage{braket}
\usepackage{color}
\usepackage{dcolumn}
\usepackage{bm}
\usepackage{physics}
\usepackage{graphicx}
\usepackage{epstopdf}
\usepackage{siunitx}

\title{
Systematic Evolution of Magnetic Anisotropy and Crystal-electric-field Ground States in Sm{\it Tr}$_2$Ge$_2$
}

\author{
Ryuji Higashinaka\thanks{E-mail address: higashin@tmu.ac.jp}, Kokoro Higo, Tatsuma D. Matsuda, and Yuji Aoki
}
\inst{Department of Physics, Tokyo Metropolitan University, Hachioji-shi, Tokyo 192-0397, Japan\\
}

\abst{
We investigated the crystal-electric-field (CEF) states in Sm{\it Tr}$_2$Ge$_2$ ({\it Tr} = Co--Cu and Ag) with the ThCr$_2$Si$_2$-type structure by magnetization and specific heat measurements.
The magnetic anisotropy evolves systematically with {\it Tr}, changing from an easy $ab$ plane for {\it Tr} = Co and Ni to an easy $c$ axis for {\it Tr} = Cu and Ag.
Combined analyses of the low-temperature susceptibility and magnetic entropy indicate a level crossing of the CEF ground state between Ni and Cu, together with a systematic evolution of the first excited CEF energy.
The evolution of the magnetic properties correlates with changes in local structural parameters, providing an experimental basis for discussing how the ligand environment around the Sm ion is related to the $4{\it f}$ electronic state.
}
\date{\today}
\begin{document}
\maketitle
\section{Introduction}
Rare-earth 4{\it f} electrons give rise to a wide variety of physical properties in solids owing to their dual nature between localization and itinerancy.
When they remain well localized, the crystalline electric field (CEF) lifts the degeneracy of the 4{\it f} multiplet, resulting in a series of discrete energy levels that govern magnetic anisotropy and multipolar degrees of freedom.
In contrast, when hybridization with conduction electrons becomes significant, itinerant heavy-fermion states characterized by strongly enhanced quasiparticle masses can emerge.
Such behavior has been extensively studied in Ce- and Yb-based compounds, and a general understanding has been established within the framework of the Kondo lattice and Fermi-liquid theory.\cite{Stewart2001,Flouquet2005}

In recent years, however, Sm-based compounds have attracted renewed attention because they exhibit more complex electronic states involving strong interplay among spin, orbital, and valence degrees of freedom.
In contrast to Ce$^{3+}$ or Yb$^{3+}$ systems, Sm ions often lie near a boundary between trivalent and intermediate-valence configurations, leading to enhanced valence fluctuations and unconventional magnetic responses.
These characteristics have been highlighted in several Sm-based intermetallic systems, including Sm{\it Tr}$_2$Al$_{20}$, where field-insensitive phase transitions and heavy-electron formation inside ordered phases have been reported.\cite{Higashinaka_2011,Yamada_2013,Yamada_2015,Sakai_2011,SmTi2Al20_24}
Experimental studies on nearly trivalent Sm compounds such as SmPt$_2$Si$_2$ and SmAu$_3$Al$_7$ have also suggested that hidden degrees of freedom remain active within magnetically ordered states.\cite{SmPt2Si2,SmPt2Si2_21,SmAu3Al7_23,SmAu3Al7_26}

To clarify the essential nature of such complex 4{\it f} electronic states in Sm systems, it is indispensable to first establish the basic properties of well-defined trivalent Sm$^{3+}$ compounds and to identify systematic trends governed by crystal field effects.
In particular, understanding how the CEF ground state and excitation spectrum evolve in a controlled manner is crucial, since these directly determine the magnetic anisotropy and low-temperature degrees of freedom.

A suitable platform for such a study is provided by intermetallic compounds with the ThCr$_2$Si$_2$-type structure (space group $I4/mmm$), which host a wide variety of correlated-electron phenomena including magnetism, heavy-fermion behavior, and unconventional superconductivity.\cite{Thompson2012}
Among them, the {\it LnTr}$_2$Ge$_2$ ({\it Ln}: lanthanide, {\it Tr}: transition metal) family has been extensively studied from structural and magnetic viewpoints, revealing systematic trends associated with the lanthanide contraction and the transition-metal substitution.\cite{SmFe2Ge2_04,SmCo2Ge2_06,SmCo2Ge2_14,SmNi2Ge2_99,SmCu2Ge2_15,SmRuRh2Ge2_96,SmAg2Ge2_09,SmIr2Ge2_85,Isikawa2014,Isikawa2016}
However, despite these efforts, detailed investigations of Sm-based members remain limited, particularly in terms of high-quality single-crystal studies and systematic comparisons across different transition metals.

In this study, we focus on the Sm{\it Tr}$_2$Ge$_2$ series ({\it Tr} = Co, Ni, Cu, and Ag) as a model system to investigate the evolution of crystal-electric-field states and magnetic anisotropy.
By magnetization and specific heat measurements on high-quality single crystals, we demonstrate a systematic change in magnetic anisotropy from easy-plane to easy-axis behavior across the series.
The analysis of the magnetic entropy further reveals a corresponding evolution of the first excited CEF energy, consistent with a change in CEF ground state.
These results provide evidence for a transition between different CEF schemes in Sm$^{3+}$ systems and provide an experimental basis for discussing how changes in the local ligand environment are related to the $4{\it f}$ electronic state.

\section{Experimental Details}
To obtain higher-quality crystals than in previous studies, the growth conditions were optimized for each compound using several fluxes.
For all flux-growth conditions, the starting materials were heated to \SI{1100}{\celsius} and kept at this temperature for 24~h before cooling.
Single crystals of Sm{\it Tr}$_2$Ge$_2$ ({\it Tr} = Co, Ni, Cu, and Ag), as well as the nonmagnetic reference compounds LaCu$_2$Ge$_2$ and LaAg$_2$Ge$_2$, were grown by flux methods, whereas polycrystalline SmPd$_2$Ge$_2$ and La{\it Tr}$_2$Ge$_2$ ({\it Tr} = Co, Ni, and Pd) were prepared by arc melting.
The detailed growth conditions are summarized in Table~\ref{Flux_List}, which shows the starting molar ratios of Sm (or La), {\it Tr}, Ge, and flux, together with the cooling rate from the maximum and decanting temperatures.
For the Cu--Ge and Ag--Ge fluxes, eutectic compositions were employed.
The obtained single crystals typically had a plate-like shape with a large (001) facet and dimensions of approximately $2 \times 2 \times 0.5$ mm$^3$.

The detailed crystal structures of Sm{\it Tr}$_2$Ge$_2$({\it Tr} = Co, Ni, Cu, and Ag) at room temperature were determined by single-crystal X-ray diffractometry (XtaLAB mini, Rigaku) with graphite monochromated Mo-K$\alpha$ radiation. 
A selected small single crystal with dimensions of roughly 0.10 $\times$ 0.08 $\times$ 0.08 mm$^3$ was mounted on a glass fiber with epoxy. 
The structural parameters of these compounds at room temperature refined using the SHELX-97 program\cite{SHELEX} are summarized in Table \ref{t1}. 

The DC magnetic susceptibility $\chi$ and the magnetization $M$ were measured in a magnetic property measurement system [MPMS; Quantum Design (QD)] down to 1.9 K and up to 7 T. 
Specific heat ($C_p$) was measured by the thermal relaxation method with a physical property measurement system (PPMS; QD) down to 2 K and up to 9 T. 
Electrical resistivity was measured on plate-like samples with a (001) facet, with the current applied along the [100] direction, using the AC transport option of the PPMS down to 2 K and in magnetic fields up to 9 T.

\section{Results and Discussion}
Figure~\ref{Tdep_Res} shows the temperature dependence of the electrical resistivity $\rho$ of Sm{\it Tr}$_2$Ge$_2$ ({\it Tr} = Co, Ni, Cu, and Ag) measured at zero field with the current along [100] up to room temperature. 
The inset displays an expanded view of the low-temperature region together with the temperature derivative $d\rho/dT$.
The residual resistivity ratio (RRR), defined as the ratio of the room-temperature resistivity to the residual resistivity, ranges from 13 to 21. 
For {\it Tr} = Co and Ni, these values are substantially higher than the previously reported values of 4.3\cite{SmCo2Ge2_14} and 6.7\cite{SmNi2Ge2_99}, demonstrating the improved quality of the present single crystals. 
The resistivity behavior below $T_{\rm N}$ can be classified into two groups. 
For {\it Tr} = Co and Ni, the decrease in resistivity is small, whereas a peak is evident in $d\rho /dT$. 
In contrast, for {\it Tr} = Cu and Ag, a clear reduction in resistivity is observed below $T_{\rm N}$, accompanied by a pronounced anomaly in $d\rho /dT$. 
Since the RRR values are comparable among these samples, the difference in low-temperature resistivity behavior is likely intrinsic and may reflect differences in magnetic states realized below $T_{\rm N}$.
For comparison, Fig.~\ref{Tdep_Res_hikaku} shows plots of all available resistivity data, including the previously reported data for {\it Tr} = Fe.\cite{SmFe2Ge2_04}
Except for {\it Tr} = Fe, the compounds show a nearly linear resistivity above 50 K, reflecting the dominant contribution from phonon scattering.
By contrast, SmFe$_2$Ge$_2$ exhibits a higher resistivity than the other compounds and a pronounced convex-upward nonlinearity in the temperature range of roughly 100--300 K.
This characteristic behavior is likely related to a substantial contribution of the Fe {\it d} band to the electrical transport.
Thus, the resistivity data provide not only a measure of sample quality through the RRR values but also information relevant to the magnetic anisotropy and the high-temperature susceptibility analysis discussed below.

%
%
Figure~\ref{Tdep_chi} shows the temperature dependence of the magnetic susceptibility of Sm{\it Tr}$_2$Ge$_2$ ({\it Tr} = Co, Ni, Cu, and Ag) measured at $H$ = 1 T for $H\ \|\ a$ and $H\ \|\ c$, together with those of the nonmagnetic reference compounds La{\it Tr}$_2$Ge$_2$ having the same crystal structure.
The inset highlights the low-temperature region near the transitions.
In all compounds, the susceptibility in the paramagnetic state just above the transition temperature can be approximately described by the Curie Weiss-like behavior over limited temperature ranges, consistent with the Sm ions being in the magnetic trivalent state Sm$^{3+}$.

Pronounced anisotropy is observed in the magnetic susceptibility, as expected from the CEF effect acting on Sm$^{3+}$ ions.
The transition temperatures and magnetic easy directions for all the compounds are shown in Table~\ref{FittingPara_all}. 
The magnetic properties can be classified into two groups.
For {\it Tr} = Co and Ni, the compounds exhibit easy-plane anisotropy within the $ab$ plane and successive magnetic transitions.
In contrast, for {\it Tr} = Cu and Ag, they show easy-axis anisotropy along the $c$-direction and only a single antiferromagnetic transition.
We use the term easy-plane anisotropy here rather than $a$-axis anisotropy, because the CEF-ground-state analysis discussed below indicates that the relevant low-energy anisotropy is associated with the distinction between the $ab$ plane and the $c$-axis, not with a particular in-plane direction.
Accordingly, the susceptibility data for Co and Ni are more naturally interpreted as reflecting an $ab$-plane-favored CEF state, whereas those for Cu and Ag reflect a $c$-axis-favored one.
For {\it Tr} = Fe, the previously reported data also show behavior similar to that of {\it Tr} = Co and Ni.\cite{SmFe2Ge2_04}
This classification is also consistent with the resistivity data, in which the reduction below $T_{\rm N}$ is small for {\it Tr} = Co and Ni but pronounced for {\it Tr} = Cu and Ag.
If the ordered state retains its easy-plane anisotropy, magnetic fluctuations within the $ab$ plane can remain active below $T_{\rm N}$ and continue to contribute to magnetic scattering, resulting in a relatively modest drop in resistivity.
By contrast, when the anisotropy is along the $c$-axis, the magnetic degrees of freedom are more strongly constrained, so the magnetic scattering is reduced more effectively below the transition.
These trends suggest that the magnetic anisotropy is closely linked to the underlying CEF ground state.
This motivates the following analysis of the CEF level scheme based on the magnetic susceptibility and entropy data.

The magnetic susceptibility of Sm compounds contains a Curie Weiss term and an approximately temperature-independent contribution, the latter including the core diamagnetism, Pauli paramagnetism, and the Van Vleck term arising from the multiplet splitting between the $J=5/2$ ground multiplet and the excited $J=7/2$ multiplet.
To estimate the Van Vleck contribution, we subtract the susceptibility of the corresponding nonmagnetic La compound from that of the Sm compound and analyze the resulting $\chi_{\rm Sm-La}$ in the temperature range of 150--300 K using
\begin{equation}
\chi_{\rm Sm-La} = \frac{C_{\rm HT}}{T-\theta_p} + \chi_{\rm VV},
\end{equation}
where $\chi_{\rm Sm-La}$ is the susceptibility difference between the Sm and La compounds, $C_{\rm HT}$ is the Curie constant in the high-temperature region, $\theta_p$ is the Weiss temperature, and $\chi_{\rm VV}$ is the Van Vleck contribution.
The Van Vleck term is expressed as
\begin{equation}
\chi_{\rm VV} = \frac{20N_a \mu_{\rm B}^2}{7k_{\rm B} \Delta_{\rm VV}},
\end{equation}
where $N_a$ is Avogadro's number, $\mu_{\rm B}$ is the Bohr magneton, $k_{\rm B}$ is the Boltzmann constant, and $\Delta_{\rm VV}$ is the energy separation between the $J=5/2$ ground multiplet and the excited $J=7/2$ multiplet.\cite{VanVleck1932} This expression is valid in the limit $\Delta_{\rm VV} \gg T$.
In this analysis, subtracting the susceptibility of the La analog removes the core diamagnetism and provides an estimate of the Pauli contribution from the non-$4f$ conduction bands, so that $\chi_{\rm Sm-La}$ mainly reflects the Curie Weiss term of localized Sm moments and the Van Vleck contribution of Sm.
This procedure is appropriate in the present compounds because the resistivity of the Sm compounds shows no clear signature of strong $c$--$f$ hybridization in the measured temperature range, so the formation of a heavy-electron state is unlikely and an additional Pauli term originating from the Sm $4f$ electrons is expected to be very small.
In addition, the convex-upward temperature dependence observed most clearly for {\it Tr} = Fe suggests a significant contribution of transition-metal $d$ states to the electronic density of states.
This point is important for interpreting the La subtraction, because an enhanced $d$-electron density of states in the nonmagnetic reference compounds is directly related to an enhanced Pauli paramagnetic contribution.
Accordingly, except for cases where the transition-metal-derived density of states changes markedly between the La and Sm compounds, the Pauli contribution can be treated to a good approximation by the La subtraction, and the remaining temperature-independent term can be analyzed phenomenologically within 150--300 K.
Because the fitted $\Delta_{\rm VV}$ values are of the 1000--1200 K order for all compounds except the Co case, the Van Vleck contribution from the excited $J=7/2$ multiplet is also expected to vary only slightly over this temperature range, and Eq.~(1) provides a reasonable description of $\chi_{\rm Sm-La}$.
The fitting results for all samples are summarized in Table~\ref{FittingPara_all}.

Among the La reference compounds, LaCo$_2$Ge$_2$ shows a much higher absolute susceptibility than the other La compounds, and the resulting $\Delta_{\rm VV}$ reaches 3000--4000 K.
This is presumably because LaCo$_2$Ge$_2$ and SmCo$_2$Ge$_2$ have different densities of states of the 3{\it d} band at the Fermi level, so that the Pauli paramagnetic contribution is not canceled sufficiently by the La subtraction and an accurate estimate of the Van Vleck susceptibility becomes difficult.
This peculiarity may reflect an enhanced contribution of Co-derived states to the magnetic response compared with the other compounds.
At present, however, we do not obtain direct experimental evidence for localized 3{\it d} magnetism or explicit $4f$--$3d$ magnetic correlations in the Co compound.
For {\it Tr} = Fe, the room-temperature susceptibilities of the nonmagnetic compounds YFe$_2$Ge$_2$, LaFe$_2$Ge$_2$, and LuFe$_2$Ge$_2$ differ markedly from one another, suggesting that the Fe 3{\it d} band contributes to the Fermi surface even more strongly than in the Co case.\cite{YFe2Ge2_16,YFe2Ge2_24,LuFe2Ge2_11,LaFe2Ge2_16}
For all compounds except SmCo$_2$Ge$_2$, $\Delta_{\rm VV}$ falls in the temperature range of 1000--1250 K.
Within the experimental uncertainty, this indicates that the energy separation between the $J=5/2$ and $J=7/2$ multiplets of Sm$^{3+}$ is nearly independent of the transition-metal species once the transition-metal contribution to the Pauli term is not anomalously large.

Next, we discuss the specific heat of Sm{\it Tr}$_2$Ge$_2$ ({\it Tr} = Co, Ni, Cu, Pd, and Ag).
The specific heat of the present compounds, $C_{\rm Sm}$, can be written as the sum of the electronic, lattice, and magnetic contributions as
\begin{equation}
C_{\rm Sm} = C_{\rm el} + C_{\rm lat} + C_{\rm mag}. 
\end{equation}
As shown in Fig.~\ref{C_S_mag}(a), we roughly estimate the magnetic contribution as $C_{\rm mag}=C_{\rm Sm}-C_{\rm La}$ by using the specific heat of the nonmagnetic isostructural compound La{\it Tr}$_2$Ge$_2$ for the electronic and lattice contributions, and we define the corresponding entropy as $S_{\rm mag}$ by integrating $C_{\rm mag}/T$ [Figs.~\ref{C_S_mag}(b)--\ref{C_S_mag}(f)].
Because the phonon spectra of the Sm and La analogs are not identical, this procedure should be regarded as an approximation, especially at elevated temperatures and for the Pd compound; however, the systematic low-temperature trends discussed below are robust against this uncertainty.
A simple Debye-model estimate for the Co compound, based on the mass dependence of the Debye temperature, gives only a small difference in lattice specific heat between SmCo$_2$Ge$_2$ and LaCo$_2$Ge$_2$ at around 30 K (see Supplemental Material\cite{SM1}), indicating that the resulting correction does not materially affect the estimate of $\Delta_{\rm CEF}$.
Consistent with the susceptibility data, for group (i), two successive transitions are observed for {\it Tr} = Co and Ni, whereas for grounp (ii), only a single transition is observed for {\it Tr} = Cu, Pd, and Ag.
The magnetic entropy also exhibits a systematic {\it Tr} dependence: near $T_{\rm N}$, it remains close to $R\ln 2$ for {\it Tr} = Co and Ag, whereas it increases markedly for {\it Tr} = Ni, Cu, and Pd, suggesting that the energy of the first excited CEF level varies across the series.

Figure~\ref{HTPD} shows the magnetic field--temperature phase diagrams constructed from the temperature dependence of magnetic susceptibility and specific heat.
For Sm{\it Tr}$_2$Ge$_2$ ({\it Tr} = Co and Ni), where two successive transitions are observed, the higher-temperature transition shows no pronounced field dependence within the field range investigated in this study.
By contrast, the lower-temperature transition shifts to lower temperatures with increasing field, indicating an intermediate ordered phase.
Among the compounds showing only a single transition, no clear field dependence of the transition temperature is observed for {\it Tr} = Cu within the measured field range, whereas the transition shifts to lower temperatures under the magnetic field for {\it Tr} = Ag.

To gain further insight into the intermediate ordered phase found for {\it Tr} = Ni, we performed detailed magnetization measurements and found that $M/H$ exhibits an inflection-like field dependence at around 3 T [Fig.~\ref{SmNi_MH}(a)].
In the intermediate phase, the field dependence is convex downward at low fields and convex upward at high fields, with the behavior changing at around 3 T.
Outside the intermediate phase, by contrast, the field dependence of $M/H$ is weak.
In trivalent Sm compounds, the response to magnetic field is generally modest because the magnetic moment of the Sm ion is small.
Nevertheless, the field response becomes pronounced in the intermediate phase of SmNi$_2$Ge$_2$.
As shown in Fig.~\ref{SmNi_MH}(b), the susceptibility also tends to increase with magnetic field in this phase.
A similar field response is observed in the intermediate phase of the Co compound.
In previous studies using lower-quality samples,\cite{SmCo2Ge2_06,SmCo2Ge2_14} disorder arising from impurities and crystal defects likely broadened the distribution of transition temperatures, causing the intermediate phase, which should be separated from the other phases, to be smeared out and thus not clearly resolved.
In this study, precise measurements using high-quality single crystals with large RRR enabled us to identify this intermediate phase clearly for the first time.

Based on the above experimental results, we now discuss the CEF splitting and energy-level scheme of Sm $4{\it f}$ electrons.
The magnetism of the present compounds originates from the Sm$^{3+}$ $4{\it f}$ electrons, and the local symmetry of the Sm site is tetragonal with point symmetry $4/mmm$.
Accordingly, the CEF Hamiltonian under tetragonal symmetry is given by
\begin{equation}
{\mathcal H}_{\rm CEF} = B_2^0 O_2^0 + B_4^0 O_4^0 + B_4^4 O_4^4 + B_6^0 O_6^0 + B_6^4 O_6^4,
\end{equation}
where $B_l^m$ and $O_l^m$ are CEF parameters and Stevens operators, respectively\cite{CEF_1, CEF_2}. 
For Sm$^{3+}$, the last two terms vanish.
Under this CEF Hamiltonian, the sixfold-degenerate $J=5/2$ multiplet is split into three Kramers doublets in tetragonal symmetry.
Their wave functions are given by
\begin{equation}
\Gamma_6 = \ket{\pm \frac{1}{2}},\ \Gamma_7^1 = \alpha\ket{\pm \frac{5}{2}} - \sqrt{1-\alpha^2}\ket{\mp \frac{3}{2}},\ \Gamma_7^2 = \sqrt{1-\alpha^2}\ket{\pm \frac{5}{2}} + \alpha\ket{\mp \frac{3}{2}}\ (0 \leqq \alpha \leqq 1). 
\end{equation} 

The magnetic moment of each eigenstate is given by $g_J \langle J_i \rangle \mu_{\rm B}$ ($i=x$, $y$, and $z$), and the expected anisotropy depends sensitively on the CEF ground state.
The calculated effective moments for representative CEF doublets are summarized in Table~\ref{mu_eff_cal}.
For an ideal $\Gamma_6$ ground state, the magnetic response is an easy plane, whereas a $\Gamma_7$ ground state near $\alpha=0$ or 1 yields an Ising-like CEF doublet with the easy axis along $c$.
In the present discussion of the magnetic anisotropy near and below $T_{\rm N}$, the relevant temperature scale is much smaller than $\Delta_{\rm VV} \sim 1000$ K, so it is sufficient to consider only the CEF splitting within the $J=5/2$ multiplet.
To identify the actual CEF ground states, we analyzed the low-temperature susceptibilities in the paramagnetic region above $T_{\rm N}$\cite{SM3}, where the contribution from higher CEF levels is minimized, and we compared the resulting effective moments with those expected for each doublet.
We also estimated the first excited CEF energy $\Delta_{\rm CEF}$ from fits to the magnetic entropy in the paramagnetic state\cite{SM4}.
Representative fits are shown in Fig.~\ref{Chi_Smag_fit}, whereas the extracted parameters are summarized in Table~\ref{FittingPara_all} and Fig.~\ref{F_FittingPara_all}.

For SmFe$_2$Ge$_2$, we used the magnetization data reported in the literature.\cite{SmFe2Ge2_04}
The low-temperature effective moments indicate that the CEF ground states are $\Gamma_6$ for group (i) and $\Gamma_7^1$ for group (ii).
For {\it Tr} = Ni and Cu, which lie near the boundary between the two groups, the effective moments deviate from the limiting values expected for isolated doublets, suggesting a substantial thermal population of the low-lying excited state.
This low-lying excited state also provides a natural explanation for the temperature evolution of the magnetic anisotropy near the boundary region.
For Sm{\it Tr}$_2$Ge$_2$ ({\it Tr} = Cu and Ag), a comparison with the calculated anisotropic moments yields the most plausible estimates $\alpha \sim 0$ and 0.09, respectively.
The detailed $\alpha$ dependence used for this estimate is provided in Fig.~S1 of Supplemental Material\cite{SM2}.
For {\it Tr} = Ag, Bala \textit{et al}. estimated $\alpha = 0.0423$,\cite{SmAg2Ge2_25} which is in reasonable agreement with our result.

The entropy analysis further shows a systematic evolution of $\Delta_{\rm CEF}$ across the series.
The full temperature dependences of $S_{\rm mag}$ for all compounds are compiled in Fig.~S2 of Supplemental Material\cite{SM3}.
For compounds far from the boundary between groups (i) and (ii), namely, {\it Tr} = Fe and Ag, the entropy tends to saturate near $R\ln 2$ around the transition temperature, indicating that the first excited CEF level lies well above the ordering scale.
By contrast, on approaching the boundary between groups (i) and (ii), the entropy exceeds $R\ln 2$ and approaches $R\ln 4$, implying that the first excited level becomes close in energy to the ground state doublet.
These observations are consistent with the fitted values of $\Delta_{\rm CEF}$.

Taken together, these results suggest that, with increasing {\it d}-electron number at the {\it Tr} site from Fe, the gap between the $\Gamma_6$ ground state and the first excited state $\Gamma_7^1$ decreases systematically, the CEF ground state changes to $\Gamma_7^1$ between Ni and Cu, and the gap then increases again.
To the best of our knowledge, this is the first study to reveal such a systematic {\it Tr} dependence of the CEF ground state in Sm compounds with the ThCr$_2$Si$_2$-type structure.

Finally, we discuss the relationship between the CEF ground state and the crystal structure.
Figure~\ref{Structure_1-2-2} shows the ${\it Ln}^{3+}$ ionic radius dependence of the unit-cell volume and lattice parameters for {\it LnTr}$_2$Ge$_2$ ({\it Tr} = Fe--Cu and Ru--Ag).\cite{LP_review, LP_01, LP_02, LP_03, LP_04, LP_05, LP_06, LP_07, LP_08, SmFe2Ge2_04, LP_10, LP_11, SmCo2Ge2_14, LP_12, LP_13, LP_14, LP_15, LP_16, LP_17, LP_18, LP_19, LP_20, LP_21, LP_22, LP_23, LP_24, LP_25, LP_26, LP_27, LP_28, LP_29, LP_30, LP_31, LP_32, LP_33, LP_34, SmIr2Ge2_85, LP_36, LP_37, LP_38, LP_39, LP_40, LP_41, LP_42, LP_43, LP_45, LP_46, LP_47}
Here, only compounds crystallizing into the same ThCr$_2$Si$_2$-type structure as the present system are included in order to exclude possible effects originating from structural differences.
The lattice parameter $a$ and the unit-cell volume $V$ show nearly monotonic changes with the ${\it Ln}^{3+}$ ionic radius, consistent with the ordinary lanthanide contraction.
In contrast, although the $c$-axis lattice parameter also varies systematically across the series, its ${\it Ln}^{3+}$ ionic radius dependence does not show a single trend.
More specifically, the slope $dc/d r_{{\it Ln}^{3+}}$ changes across the series, indicating that the structural evolution along the $c$-direction is qualitatively modified around the boundary where the CEF ground state changes from $\Gamma_6$ to $\Gamma_7^1$.
This correspondence indicates an experimental correlation between the changes in magnetic anisotropy and CEF ground state and the anomalous structural evolution along the $c$-axis.

To discuss this trend at a qualitative level, we examined the Ge internal coordinate $z_{\rm Ge}$.
The single-crystal structural analysis indicates that this change in slope is accompanied by a systematic variation in $z_{\rm Ge}$.
With increasing $d$-electron number from Co and Ni to Cu, Pd, and Ag, the transition-metal-derived $d$ bands are progressively filled.
Because these states are also coupled to the Ge $p$ orbitals through chemical bonding, their spatial distribution and bonding characteristics are expected to evolve across the series.
The Ge position should then be determined so as to minimize the total crystal energy under the corresponding electronic structure, and this evolution is experimentally observed as changes in $z_{\rm Ge}$ and in the derived distances $d_{\rm Ge-Ge}$ and $d_{\rm Sm-Ge}$.
Here, $d_{\rm Ge-Ge}$ denotes the interlayer Ge--Ge distance between adjacent Ge sheets along the $c$-axis, whereas $d_{\rm Sm-Ge}$ is defined as the $c$-axis distance between a Sm ion and the Ge ion located at the same $xy$ position as shown in Fig.~\ref{zGe}(a).
As shown in Figs.~\ref{zGe}(b)--\ref{zGe}(g), the variation in $z_{\rm Ge}$ does not change the local point symmetry of the Sm site, which remains tetragonal throughout the series, but it modifies the two distances $d_{\rm Ge-Ge}$ and $d_{\rm Sm-Ge}$ defined above. 
From this viewpoint, $z_{\rm Ge}$ and the associated distances $d_{\rm Ge-Ge}$ and $d_{\rm Sm-Ge}$ are useful experimental quantities for tracking the underlying evolution of the transition-metal $d$ and Ge $p$ electronic states. 
More specifically, near the boundary where $dc/d r_{{\it Ln}^{3+}}$ changes its sign from positive to negative, $z_{\rm Ge}$ increases, accompanied by a decrease in $d_{\rm Ge-Ge}$ and an increase in $d_{\rm Sm-Ge}$. 
This means that the Ge layers move closer to the $ab$ plane containing the Sm ions, whereas the Ge ions located directly above and below a given Sm site along the $c$ direction moves farther away. 
Such a geometrical change is expected to modify the space surrounding the Sm ion and thereby determine whether a more planar or uniaxial $4f$ charge distribution is favored. In this sense, the correspondence between this trend and the observed evolution from easy-plane behavior for Co and Ni to easy-axis behavior for Cu and Ag is indicated, but we regard it as a qualitative interpretation only. 
A quantitative discussion of the element-dependent CEF anisotropy would require a more microscopic evaluation of how the local potential acting on Sm evolves, for example, by first-principles calculations, and is left for future work. 
The actual evolution of the CEF scheme should therefore be viewed as reflecting coupled changes in the transition-metal $d$ bands, the Ge $p$ states, and the resulting local structure around Sm, and the present structural discussion is intended only to highlight experimentally identified quantities that track that evolution.

\section{Summary}
In this study, we systematically investigated the physical properties of single-crystal Sm${\it Tr}_2$Ge$_2$ (${\it Tr} =$ Co--Cu, Ag) with the ThCr$_2$Si$_2$-type structure.
As a result, a pronounced evolution of magnetic anisotropy was found with the change in ${\it Tr}$: the Co- and Ni-based compounds exhibit easy-plane anisotropy, whereas the Cu- and Ag-based compounds show easy-axis anisotropy along the $c$-direction.
Combined analyses of the low-temperature susceptibility and magnetic entropy indicate that the crystal-electric-field level scheme of Sm$^{3+}$ evolves systematically with ${\it Tr}$: the first excited CEF energy decreases toward the Ni--Cu boundary, the CEF ground state changes from $\Gamma_6$ to $\Gamma_7^1$ there, and the gap then increases again on the easy-axis side.
The boundary at which this interchange occurs corresponds to a change in the ${\it Ln}^{3+}$ ionic radius dependence of the $c$-axis lattice parameter, namely, a change in the slope $dc/d r_{{\it Ln}^{3+}}$.
We conclude that the evolution of the local structure, reflected in the Ge $z$ parameter and the associated changes in Sm--Ge and Ge--Ge distances, provides experimentally identified quantities that track the evolution of the transition-metal $d$ bands, the Ge $p$ states, and the resulting changes in $4{\it f}$ ground state and magnetic anisotropy. 
A quantitative microscopic account of this coupled evolution remains a subject for future work.

\begin{acknowledgments}
This work was supported by MEXT/JSPS KAKENHI Grants Number JP25K07228, JP23H04870, and JP22K03517.
\end{acknowledgments}

\clearpage
\begin{figure}
\caption{(Color online) 
Temperature dependence of the electrical resistivity of Sm{\it Tr}$_2$Ge$_2$ ({\it Tr} = Co, Ni, Cu, and Ag) at $H$ = 0 T.
The inset shows an expanded view of the low-temperature region together with the temperature derivative $d\rho/dT$.
}
\label{Tdep_Res}
\end{figure}
\begin{figure}
\caption{
Comparison of all available data on the temperature dependence of the electrical resistivity of Sm{\it Tr}$_2$Ge$_2$ ({\it Tr} = Fe, Co, Ni, Cu, and Ag) at $H$ = 0 T.
The data for {\it Tr} = Fe are taken from Avila {\it et al}.\cite{SmFe2Ge2_04}
}
\label{Tdep_Res_hikaku}
\end{figure}
\begin{figure}
\caption{(Color online) 
Temperature dependence of the magnetic susceptibility of Sm{\it Tr}$_2$Ge$_2$ ({\it Tr} = Co, Ni, Cu, and Ag) measured at $H$ = 1 T for $H\ \|\ a$ and $H\ \|\ c$.
The data for the nonmagnetic reference compounds La{\it Tr}$_2$Ge$_2$ are also shown.
The inset shows an expanded view of the low-temperature region of the susceptibility along the easy plane or direction.
}
\label{Tdep_chi}
\end{figure}
\begin{figure}
\caption{(Color online) 
(a) Temperature dependence of the specific heat of SmCo$_2$Ge$_2$.
The magnetic contribution is estimated by subtracting the specific heat of the nonmagnetic reference compound LaCo$_2$Ge$_2$.
(b)--(f) Temperature dependence of the magnetic specific heat and magnetic entropy of Sm{\it Tr}$_2$Ge$_2$ ({\it Tr} = Co, Ni, Cu, Pd, and Ag) at 0 T.
}
\label{C_S_mag}
\end{figure}
\begin{figure}
\caption{
Magnetic field--temperature phase diagrams of Sm{\it Tr}$_2$Ge$_2$ for (a) Co, (b) Ni, (c) Cu, and (d) Ag, for fields applied along the magnetic easy plane or direction.
}
\label{HTPD}
\end{figure}
\begin{figure}
\caption{
(a) Magnetization curves and (b) temperature dependence of the magnetic susceptibility in the intermediate ordered phase of SmNi$_2$Ge$_2$.
}
\label{SmNi_MH}
\end{figure}
\begin{figure}
\centering
  \caption{(Color online)
    (a) Temperature dependence of the magnetic susceptibility of SmCo$_2$Ge$_2$ at $H$ = 1 T for $H\parallel a$ and $H\parallel c$.
  The fit to the high-temperature region using Eq.~(1) and the low-temperature Curie Weiss fit are also shown.
  (b) Temperature dependence of $S_{\rm mag}$ of SmAg$_2$Ge$_2$ at 0 T.
  The fit above $T_{\rm N}$ is also shown.
}
   \label{Chi_Smag_fit}
\end{figure}
\begin{figure}
\centering
  \caption{(Color online) 
  (a, d) Effective magnetic moments for $H\parallel a$ and $H\parallel c$ obtained from Curie Weiss fits to the low-temperature susceptibility of Sm{\it Tr}$_2$Ge$_2$ ({\it Tr} = Fe--Cu, Pd, and Ag).
  The red region corresponds to $\mu_{\rm eff}^{\perp c}$ and $\mu_{\rm eff}^{\parallel c}$ expected for the $\ket{\pm \frac{1}{2}}$ ground state.
  The blue region indicates the case of the $\ket{\pm \frac{3}{2}}$ ground state.
  (b, e) First excited CEF energies $\Delta_{\rm CEF}$ obtained from the analysis of the magnetic entropy.
(c, f) Slope $dc/dr_{{\it Ln}^{3+}}$ of the lattice parameter $c$ with respect to the ${\it Ln}^{3+}$ ionic radius.
}
   \label{F_FittingPara_all}
\end{figure}
\begin{figure}
\caption{
${\it Ln}^{3+}$ ionic radius dependence of (a, d) the unit-cell volume $V$ and (b, e) lattice parameters $a$ and (c, f) $c$ for {\it LnTr}$_2$Ge$_2$.\cite{LP_review, LP_01, LP_02, LP_03, LP_04, LP_05, LP_06, LP_07, LP_08, SmFe2Ge2_04, LP_10, LP_11, SmCo2Ge2_14, LP_12, LP_13, LP_14, LP_15, LP_16, LP_17, LP_18, LP_19, LP_20, LP_21, LP_22, LP_23, LP_24, LP_25, LP_26, LP_27, LP_28, LP_29, LP_30, LP_31, LP_32, LP_33, LP_34, SmIr2Ge2_85, LP_36, LP_37, LP_38, LP_39, LP_40, LP_41, LP_42, LP_43, LP_45, LP_46, LP_47}
Because the data are plotted against trivalent ionic radii, several Eu and Yb compounds, which are suggested to take the divalent state, deviate from the systematic trend.
Solid lines are drawn as guides to the eye for compound groups in which the lattice parameter $c$ exhibits behavior opposite to the ordinary lanthanide contraction.
}
\label{Structure_1-2-2}
\end{figure}
\begin{figure}[hp]
\caption{(Color online) 
(a) Crystal structure of Sm{\it Tr}$_2$Ge$_2$, drawn using the VESTA program. \cite{VESTA}
{\it Tr}-element dependence of (b, e) the Ge internal coordinate $z_{\rm Ge}$ and the geometrical quantities derived from $z_{\rm Ge}$ and the lattice parameter $c$ for Sm{\it Tr}$_2$Ge$_2$: (c, f) the interlayer Ge--Ge distance $d_{\rm Ge-Ge} = (1-2z_{\rm Ge})c$ and (d, g) the Sm--Ge distance at the same $xy$ position, $d_{\rm Sm-Ge} = z_{\rm Ge}c$.
The corresponding distances are indicated in (a).
}
\label{zGe}
\end{figure}
\clearpage
\begin{table*}[tb]
\caption{
Crystal growth conditions for single-crystal Sm{\it Tr}$_2$Ge$_2$ ({\it Tr} = Co, Ni, Cu, and Ag) and La{\it Tr}$_2$Ge$_2$ ({\it Tr} = Cu and Ag) prepared by the flux method.
The table shows the starting molar ratios of Sm (or La), {\it Tr}, Ge, and flux before growth, together with the cooling rate and decanting temperatures.
For the Cu--Ge and Ag--Ge fluxes, eutectic compositions were used, namely, Cu:Ge = 63:37 and Ag:Ge = 75.5:24.5, with eutectic temperatures of \SI{644}{\celsius} and \SI{651}{\celsius}, respectively.
}
\begin{center}
\begin{tabular}{cccccccc}
 & \multicolumn{4}{c}{Starting molar ratio} & & Cooling & Decanting \\
 \cline{2-5} 
 {\it Tr} & Sm & {\it Tr} & Ge & Flux & & rate (\si{\celsius}/h) & temp. (\si{\celsius}) \\
 \hline \hline
 Co & 1 & 2 & 2 & 30 & Sn & 3.3 & 500 \\
 Ni & 1 & 2 & 2 & 30 & Bi & 4.5 & 500 \\
 Cu & 1 & 2 & 2 & 30 & Cu--Ge & 2.1 & 700 \\
 Ag & 1 & 2 & 2 & 19 & Ag--Ge & 2.1 & 700 \\
 \hline
 {\it Tr} & La & {\it Tr} & Ge & Flux & &  &  \\
\hline \hline
 Cu & 1 & 2.3 & 2 & 30 & In & 2.8 & 300 \\
Ag & 1 & 2 & 2 & 19 & Ag--Ge & 2.1 & 700 \\
\end{tabular}
\end{center}
\label{Flux_List}
\end{table*}
\begin{table*}[hp]
	\caption{Atomic coordinates and thermal parameters of Sm{\it Tr}$_2$Ge$_2$ ({\it Tr} = Co, Ni, Cu, Ag) at room temperature determined by single-crystal X-ray measurements. 
		The number of formula unit cell per unit cell is $Z$ = 2.
		$B_{\rm eq}$ is the equivalent isotropic atomic displacement parameter. 
		$R_1$ and $wR_2$ are reliability factors. 
		Standard deviations in the positions of the least significant digits are given in parentheses.}
	\label{t1}
	\begin{center}
		\begin{tabular}{llccclllll}
			\multicolumn{6}{l}{(a) SmCo$_2$Ge$_2$, ($a$ = 4.008(9) \AA, $c$ = 10.14(2) \AA, $V$ = 162.9(6)~\AA$^3$)} \\ \hline 
			\multicolumn{2}{l}{$I4/mmm$, $D_{4h}^{17}$  (\#139)}&  \multicolumn{3}{c}{Position} \\
			\cline{3-5}
			Atom & Site & $x$ & $y$ & $z$ & $B_{\rm eq}\ {\rm (\AA}^2)$ \\
			\hline
			Ge    & 4$e$\ ($4mm$)      & 0      & 0      & 0.37220(2)   & 0.55(6) \\
			Co & 4$d$\ (${\bar 4}m2$)      & 0      & 1/2     & 1/4   & 0.59(7) \\
			Sm & 2$a$\ ($4/mmm$)     & 0 & 0 & 0 & 0.56(6)  \\ \hline
			\hline
			& & & \multicolumn{3}{c}{$R_1$ $=$ 3.59 $\%$, $wR_2$ $=$ 8.64 $\%$} \\
			\\  
			\multicolumn{6}{l}{(b) SmNi$_2$Ge$_2$, ($a$ = 4.098(8) \AA,    $c$ = 9.84(2) \AA,   $V$ = 165.2(6)~\AA$^3$) } \\ \hline 
			\multicolumn{2}{l}{$I4/mmm$, $D_{4h}^{17}$  (\#139)}&  \multicolumn{3}{c}{Position} \\
			\cline{3-5}
			Atom & Site & $x$ & $y$ & $z$ & $B_{\rm eq}\ {\rm (\AA}^2)$ \\
			\hline
			Ge    & 4$e$\ ($4mm$)       & 0      & 0      & 0.37035(2)   & 0.54(6) \\
			Ni & 4$d$\ (${\bar 4}m2$)      & 0      & 1/2     & 1/4   & 0.54(5) \\
			Sm & 2$a$\ ($4/mmm$)      & 0 & 0 & 0 & 0.41(4)  \\ \hline
			\hline
			& & & \multicolumn{3}{c}{$R_1$ $=$ 2.00 $\%$, $wR_2$ $=$ 4.65 $\%$} \\
			\\ 
			\multicolumn{6}{l}{(c) SmCu$_2$Ge$_2$, ($a$ = 4.097(8) \AA,    $c$ = 10.25(2) \AA,   $V$ = 172.1(6)~\AA$^3$) } \\ \hline 
			\multicolumn{2}{l}{$I4/mmm$, $D_{4h}^{17}$  (\#139)}&  \multicolumn{3}{c}{Position} \\
			\cline{3-5}
			Atom & Site & $x$ & $y$ & $z$ & $B_{\rm eq}\ {\rm (\AA}^2)$ \\
			\hline
			Ge    & 4$e$\ ($4mm$)      & 0      & 0      & 0.37966(10)   & 0.59(7) \\
			Cu & 4$d$\ (${\bar 4}m2$)      & 0      & 1/2     & 1/4   & 0.73(6) \\
			Sm & 2$a$\ ($4/mmm$)      & 0 & 0 & 0 & 0.50(5)  \\ \hline
			\hline
			& & & \multicolumn{3}{c}{$R_1$ $=$ 2.41$\%$, $wR_2$ $=$ 5.92 $\%$} \\
			\\ 
			\multicolumn{6}{l}{(d) SmAg$_2$Ge$_2$, ($a$ = 4.2308(11) \AA,    $c$ = 11.055(3) \AA,   $V$ = 167.88(9)~\AA$^3$) } \\ \hline 
			\multicolumn{2}{l}{$I4/mmm$, $D_{4h}^{17}$  (\#139)}&  \multicolumn{3}{c}{Position} \\
			\cline{3-5}
			Atom & Site & $x$ & $y$ & $z$ & $B_{\rm eq}\ {\rm (\AA}^2)$ \\
			\hline
			Ge    & 4$e$\ ($4mm$)       & 0      & 0      & 0.39126(13)   & 0.84(4) \\
			Ag & 4$d$\ (${\bar 4}m2$)     & 0      & 1/2     & 1/4   & 0.89(4) \\
			Sm & 2$a$\ ($4/mmm$)     & 0 & 0 & 0 & 0.62(4)  \\ \hline
			\hline
			& & & \multicolumn{3}{c}{$R_1$ $=$ 2.18 $\%$, $wR_2$ $=$ 5.07 $\%$} \\
			\\ 
		\end{tabular}
	\end{center}
\end{table*}
\begin{table*}[hp]
\caption{
Magnetic transition temperatures, magnetic easy directions, energy gap between the $J=5/2$ ground multiplet and the excited $J=7/2$ multiplet $\Delta_{\rm VV}$, low-temperature effective magnetic moments $\mu_{\rm eff}$, inferred CEF ground states, and first excited CEF energies $\Delta_{\rm CEF}$ for Sm{\it Tr}$_2$Ge$_2$ ({\it Tr} = Fe--Cu, Pd, and Ag).
For {\it Tr} = Pd, the anisotropy is presently unknown because only polycrystalline data are available.

}
\centering
\begin{tabular}{cccccccccc}
\hline \hline
 & & magnetic easy & \multicolumn{2}{c}{$\Delta_{\rm VV}\ ({\rm K})$}  & & \multicolumn{2}{c}{$\mu_{\rm eff}$ ($\mu_{\rm B}$)} & CEF & \\
\cline{4-5} \cline{7-8}
{\it Tr} &  $T_{\rm N}$ (K) &direction & $H\ \|\ a$ & $H\ \|\ c$ & & $H\ \|\ a$ & $H\ \|\ c$ & ground state & $\Delta_{\rm CEF}$ (K)\\
\hline \hline
Fe & 5.9, 4.9\cite{SmFe2Ge2_04} & $ab$ plane & & & & 0.760 & 0.380 & $\Gamma_6$ & -- \\
Co & 18.6, 16.5 & $ab$ plane & 4070 & 2960 & & 0.747 & 0.311 & $\Gamma_6$ & 113 \\
Ni & 19.6, 15.8 & $ab$ plane & 1240 & 1020 & & 0.749 & 0.522 & $\Gamma_6$ & 40 \\
Cu & 15.0 & $c$ axis & 1120 & 1200 & & $\approx$ 0 & 0.839 & $\Gamma_7^{1} \; (\alpha \approx 0)$ & 36\\
\hline 
Pd & 4.8 & -- & -- & -- & & -- & -- & -- & 13 \\
Ag & 9.2 & $c$ axis & 1040 & 1060 & & 0.103 & 0.576 & $\Gamma_7^{1} \; (\alpha \approx 0.09)$ & 106 \\
\hline \hline
\end{tabular}
\label{FittingPara_all}
\end{table*}
\begin{table*}[tb]
\caption{
Calculated effective magnetic moments of representative Sm$^{3+}$ CEF doublets under tetragonal symmetry for field components parallel and perpendicular to the $c$-axis.
}
\begin{center}
\begin{tabular}{lcc}
 & \multicolumn{2}{c}{$\mu_{\rm eff}$ ($\mu_{\rm B}$)} \\
 \cline{2-3}
 & $\mu_{\rm eff}^{\| c}$ & $\mu_{\rm eff}^{\perp c}$ \\
 \hline \hline
$\Gamma_6$ (= $\ket{\pm \frac{1}{2}}$) & 0.247 & 0.742 \\
$\Gamma_7^1 (\alpha = 0)$ (= $\ket{\pm \frac{3}{2}}$) & 0.742 & 0 \\
$\Gamma_7^1 (\alpha = 1)$ (= $\ket{\pm \frac{5}{2}}$) & 1.237 & 0 \\
\hline \hline
\end{tabular}
\end{center}
\label{mu_eff_cal}
\end{table*}

\clearpage

\setcounter{figure}{0}
\begin{figure}
\includegraphics[width=\linewidth]{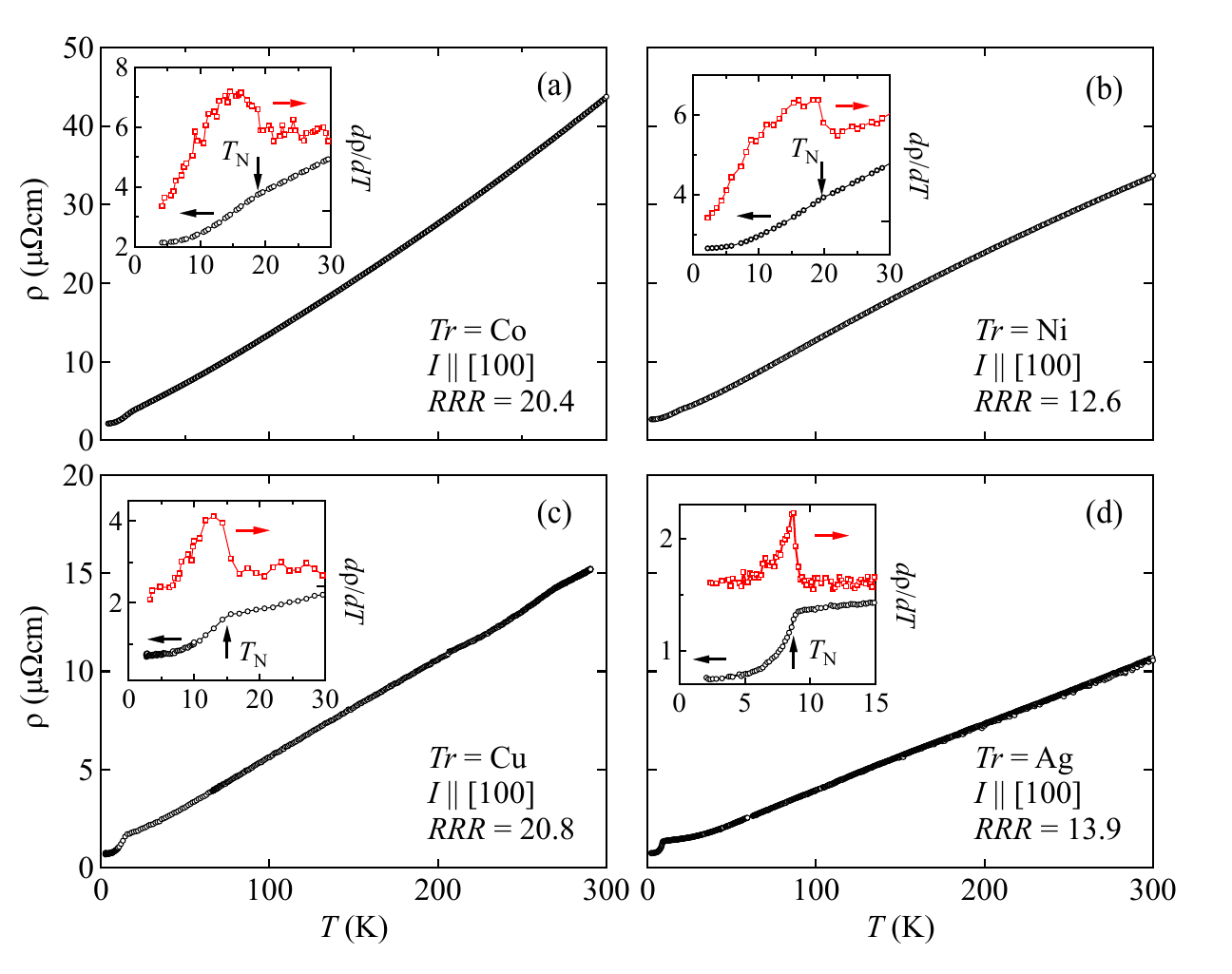}
\caption{}
\label{Tdep_Res}
\end{figure}
\begin{figure}
\includegraphics[width=\linewidth]{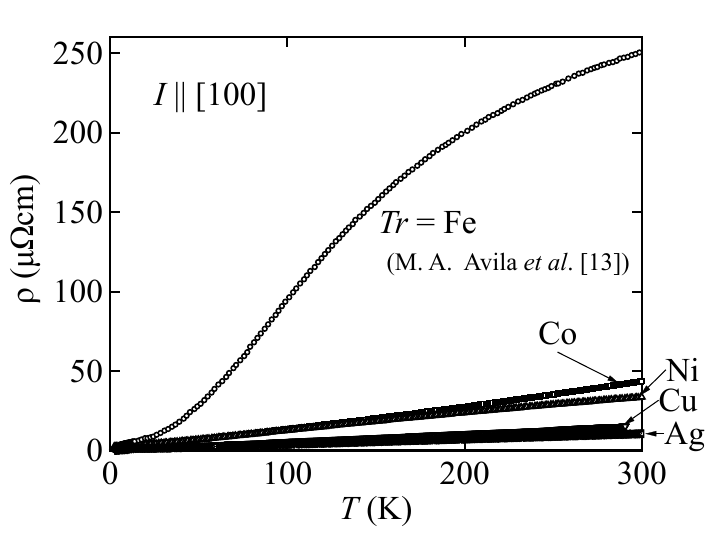}
\caption{}
\label{Tdep_Res_hikaku}
\end{figure}
\begin{figure}
\includegraphics[width=\linewidth]{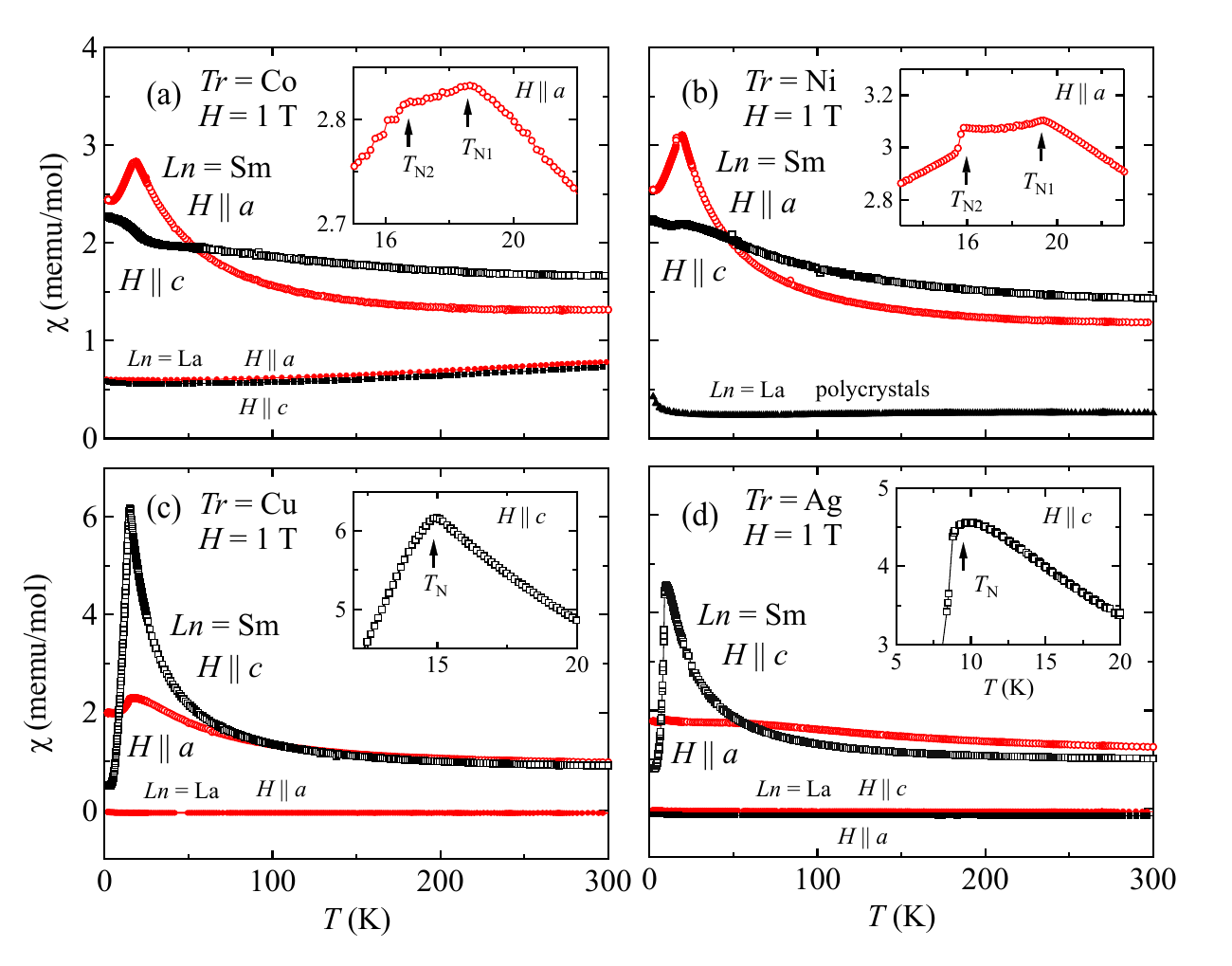}
\caption{}
\label{Tdep_chi}
\end{figure}
\begin{figure}
\includegraphics[width=\linewidth]{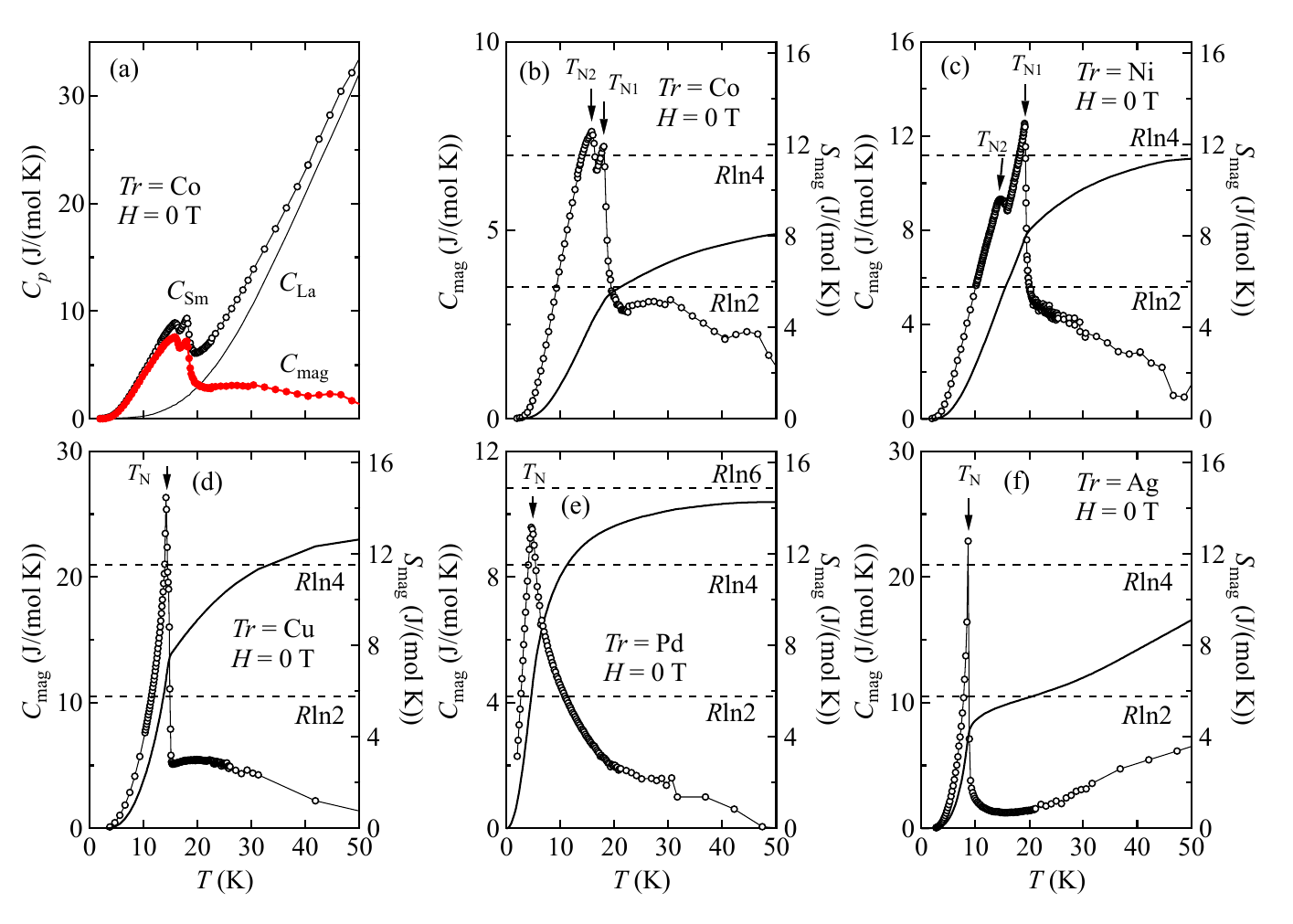}
\caption{}
\label{C_S_mag}
\end{figure}
\begin{figure}
\includegraphics[width=\linewidth]{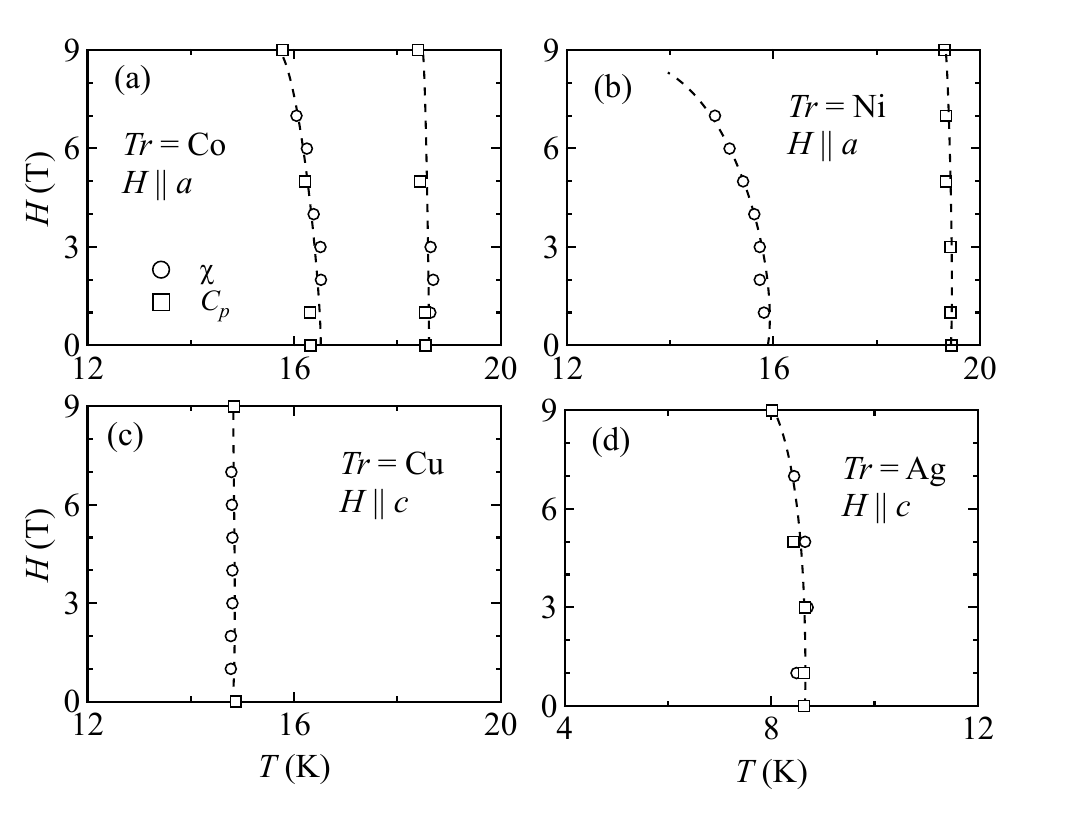}
\caption{}
\label{HTPD}
\end{figure}
\begin{figure}
\includegraphics[width=\linewidth]{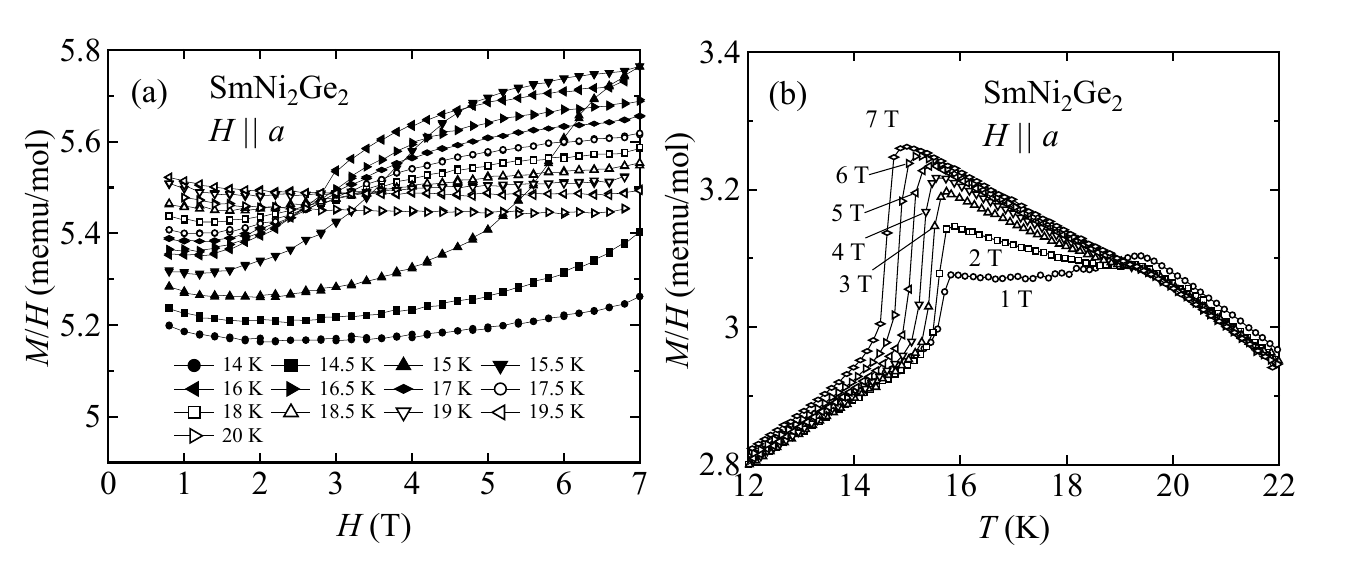}
\caption{}
\label{SmNi_MH}
\end{figure}
\begin{figure}
\includegraphics[width=\linewidth]{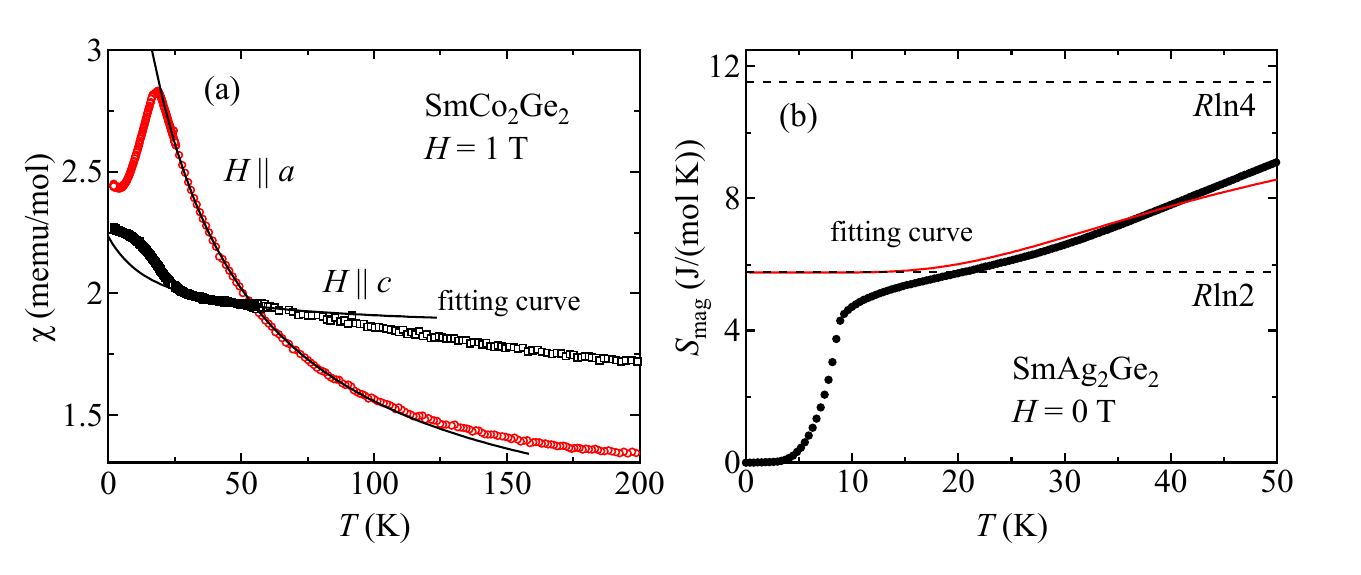}
\caption{}
\label{Chi_Smag_fit}
\end{figure}
\clearpage
\begin{figure}
\includegraphics[width=\linewidth]{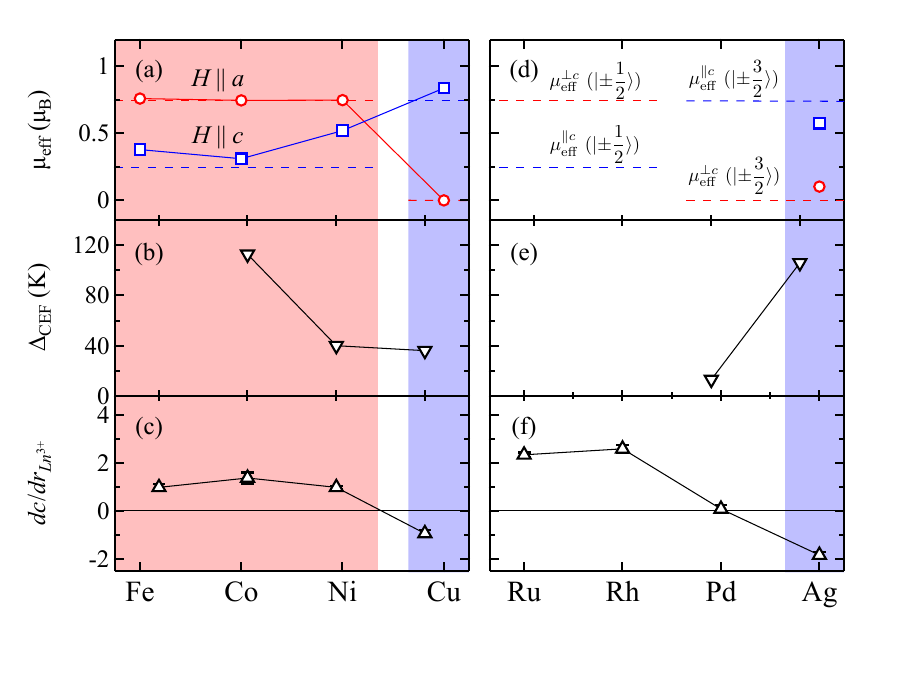}
\caption{}
\label{F_FittingPara_all}
\end{figure}
\begin{figure}
\includegraphics[width=\linewidth]{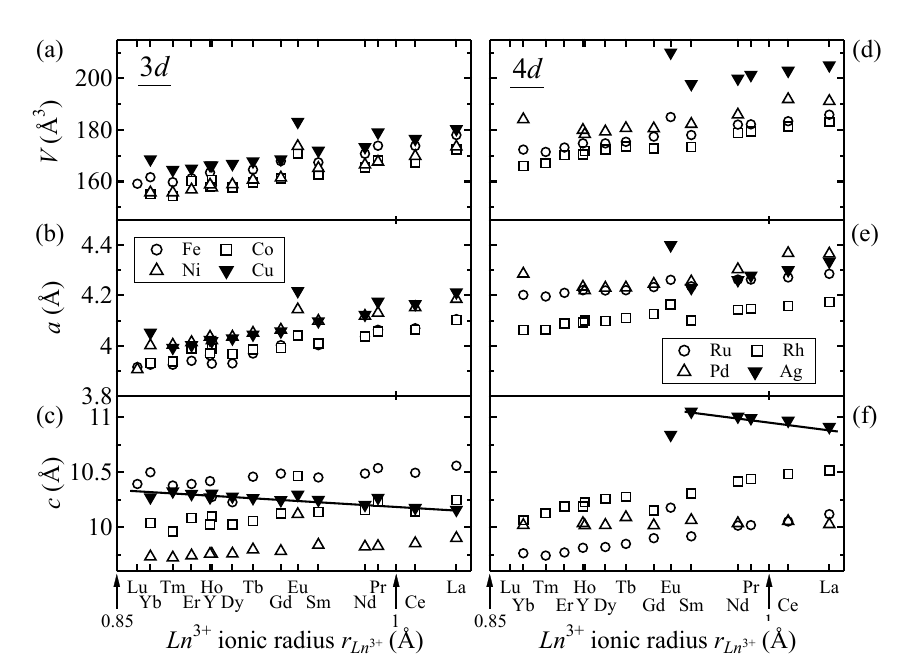}
\caption{}
\label{Structure_1-2-2}
\end{figure}
\begin{figure}[hp]
\includegraphics[width=\linewidth]{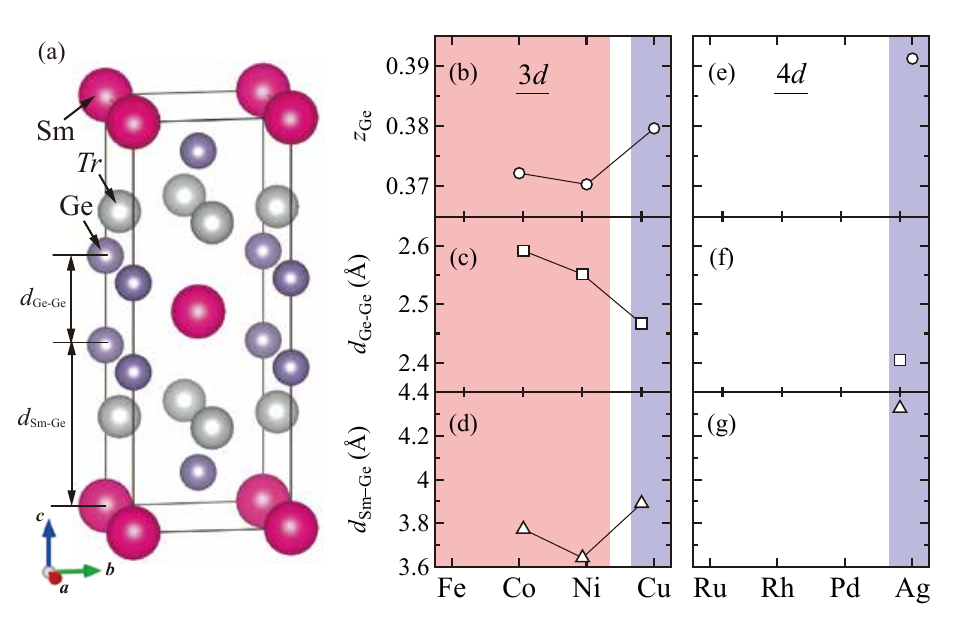}
\caption{}
\label{zGe}
\end{figure}

\end{document}